\documentclass[12pt,preprint]{aastex}
\usepackage{amsmath}
\usepackage{graphicx}
\usepackage{color}
\usepackage{url}
\usepackage{CJKutf8}
\usepackage{ulem}

\slugcomment{The Astronomical Journal}

\shorttitle{2016 J1}
\shortauthors{Hui et al.}

\begin{document}

\title{Split Active Asteroid P/2016 J1 (PANSTARRS)}
\author{
Man-To Hui
\begin{CJK}{UTF8}{bsmi}
(許文韜)$^{1}$ 
\end{CJK}
, David Jewitt$^{1,2}$
, and 
Xinnan Du 
\begin{CJK}{UTF8}{bsmi}
(杜辛楠)$^{2}$ 
\end{CJK}
}
\affil{$^1$Department of Earth, Planetary, and Space Sciences,
UCLA, 
595 Charles Young Drive East, Box 951567, 
Los Angeles, CA 90095-1567\\
}

\affil{$^2$Department of Physics and Astronomy, UCLA, 
430 Portola Plaza, Box 951547, Los Angeles, CA 90095-1547\\
}
\email{pachacoti@ucla.edu}

\begin{abstract}

We present a photometric and astrometric study of the split active asteroid P/2016 J1 (PANSTARRS). The two components (hereafter J1-A and J1-B) separated either $\sim$1500 days (2012 May to June) or 2300 days (2010 April) prior to the current epoch, with a separation speed $V_{\mathrm{sep}} = 0.70 \pm 0.02$ m s$^{-1}$ for the former scenario, or $0.83 \pm 0.06$ m s$^{-1}$ for the latter.  Keck photometry reveals that the two fragments have similar, Sun-like colors which are comparable to the colors of primitive C- and G-type asteroids. With a nominal comet-like albedo, $p_{R} = 0.04$, the effective, dust-contaminated cross sections are estimated to be 2.4 km$^{2}$ for J1-A, and 0.5 km$^{2}$ for J1-B. We estimate that the nucleus radii lie in the range $140 \lesssim R_{\mathrm{N}} \lesssim 900$ m for J1-A and $40 \lesssim R_{\mathrm{N}} \lesssim 400$ m, for J1-B. A syndyne-synchrone simulation shows that both components have been active for 3 to 6 months, by ejecting dust grains at speeds $\sim$0.5 m s$^{-1}$ with rates $\sim$1 kg s$^{-1}$ for J1-A and 0.1 kg s$^{-1}$ for J1-B. In its present orbit, the rotational spin-up and devolatilization times of 2016 J1 are very small compared to the age of the solar system, raising the question of why this object still exists. We suggest that ice that was formerly buried within this asteroid became exposed at the surface, perhaps via a small impact, and that sublimation torques then rapidly drove it to break-up. Further disintegration events are anticipated due to the rotational instability.
  
\end{abstract}

\keywords{
minor planets, asteroids: general --- minor planets, asteroids: individual (P/2016 J1) --- methods: data analysis
}

\section{\uppercase{Introduction}}

Active asteroids are solar system bodies that are dynamically asteroidal (Tisserand parameter with respect to Jupiter $T_\mathrm{J} \gtrsim 3$) but have been observed to show comet-like mass-loss activity. To date, over twenty such objects have been discovered. Remarkably there is a broad range of different physical mechanisms which account for this activity, including rotational instability, sublimation of volatiles, impacts and thermal fracture (Jewitt et al. 2015a).

P/2016 J1 (PANSTARRS; hereafter ``2016 J1")  was discovered  on UT 2016 May 5 by the Panoramic Survey Telescope and Rapid Response System (PANSTARRS). Besides its asteroidal orbit and cometary appearance, it was observed to be split into two pieces, the brighter one  designated as component J1-A  and the fainter as component J1-B by the Minor Planet Center (MPC; Wainscoat et al. 2016). Previous unambiguous discoveries of split active asteroids include P/2013 R3 (Catalina-PANSTARRS with at least 10 components; Jewitt et al. 2014a) and 331P/2012 F5 (Gibbs with at least 5 components; Drahus et al. 2015). Therefore, 2016 J1 is the third member of this type. The two components are moving in nearly identical orbits (semimajor axis $a = 3.172$ AU, eccentricity $e = 0.23$, and inclination $i = 14\degr.3$), and have an indistinguishable $T_\mathrm{J} = 3.11$.

In this paper, we analyze Keck 10 m telescope images of 2016 J1 obtained on UT 2016 August 4 for physical properties of the two components (Sections \ref{sect_phot} and \ref{sect_morph}), and we study the fragmentation dynamics based on new and published astrometric data (Section \ref{sect_sd}).

\section{\uppercase{Observations}}
\label{sect_obs}

We obtained a sequence of images  in the broadband $B$, $V$ and $R$ filters on UT 2016 August 4, using the Keck I 10 m telescope on Mauna Kea, Hawaii. At this time, 2016 J1 was at heliocentric distance $r_H$ = 2.458 AU, geocentric distance $\Delta$ = 2.065 AU and had phase angle $\alpha$ = 23.9\degr. The $B$-band observations along with the $V$- and $R$-band observations were performed simultaneously through the Low Resolution Imaging Spectrograph (LRIS), which has  independent blue and red channels separated by a dichroic beam splitter (Oke et al. 1995). We used the ``460" dichroic, which has 50\% transmission at 4875 \AA. On the blue channel, we used a $B$-band filter with effective wavelength $\lambda_\mathrm{eff} = 4370$ \AA~and full-width half maximum (FWHM) of $\Delta \lambda = 878$ \AA. On the red side, we took $V$- and $R$-band images. The $V$-band filter has $\lambda_\mathrm{eff} = 5473$ \AA~and FWHM $\Delta \lambda = 948$ \AA, and the $R$ filter has $\lambda_\mathrm{eff} = 6417$ \AA~and FWHM $\Delta \lambda = 1185$ \AA.\footnote{\url{https://www2.keck.hawaii.edu/inst/lris/filters.html}} Image exposure times were 250 s for the $B$-band data, and 200 s for both the $V$- and $R$-band data. All images were taken with the telescope tracked non-sidereally to follow the apparent motion of 2016 J1 (30\arcsec~hr$^{-1}$ in Right Ascension, $-13\arcsec$ hr$^{-1}$ in Declination, approximately). The data  were  flat-fielded using images of a diffusely illuminated patch on the inside of the Keck dome and photometric calibration was obtained from images of star PG 1633+099A from the Landolt  catalogue (Landolt 1992). The image scale of both channels is 0\arcsec.135 pixel$^{-1}$, while the seeing varied between $\sim$0\arcsec.6--0\arcsec.9 FWHM.

\section{\uppercase{Photometric Results}}
\label{sect_phot}

We used synthetic circular apertures to extract photometry from the aligned, co-added images of 2016 J1. Our photometric aperture is 2\arcsec.0 in radius. Sky values were obtained from a contiguous annulus of 2\arcsec.0 in width centered at the target. Photometric uncertainties on 2016 J1, denoted as $\sigma_{m, \lambda}$, were estimated from error propagation, i.e., $\sigma_{m, \lambda} = \sqrt{\sigma_{\ast,\mathrm{s}}^2 + \sigma_{\ast,\lambda}^2 + \sigma_\mathrm{I}^2}$, where $\sigma_{\ast,\mathrm{s}}$ is the standard deviation on the mean of repeated measurements of PG 1633+099A, $\sigma_{\ast,\lambda}$ is the magnitude error of the star in the corresponding bandpass given by Landolt (1992), and $\sigma_\mathrm{I}$ is the uncertainty in instrumental magnitudes of the two components calculated from the gain and read noise of the instrument.

For convenience we denote the apparent magnitudes of components of 2016 J1 as $m_{\lambda,j}$, where $\lambda$ refers to the bandpasses ($B$, $V$ or $R$), and $j=\mathrm{a,b}$ labelling J1-A and J1-B, respectively. Our measurements yield $m_{\mathrm{a},B} = 23.04 \pm 0.03$, $m_{\mathrm{a},V} = 22.30 \pm 0.03$, and $m_{\mathrm{a},R} = 21.94 \pm 0.03$ for   J1-A, and $m_{\mathrm{b},B} = 24.78 \pm 0.08$, $m_{\mathrm{b},V} = 24.04 \pm 0.09$, and $m_{\mathrm{b},R} = 23.65 \pm 0.08$ for J1-B. Thus we can find that two fragments have colors nearly identical to each other within the noise level: $\left( m_{B} - m_{V} \right)_{\mathrm{a}} = 0.74 \pm 0.04$ and $\left( m_V - m_R \right)_{\mathrm{a}} = 0.36 \pm 0.04$ for J1-A, and $\left( m_B - m_V \right)_{\mathrm{b}} = 0.74 \pm 0.12$ and $\left(m_V - m_R \right)_{\mathrm{b}} = 0.39 \pm 0.12$ for  J1-B, in comparison with the color indices of the Sun $\left( m_B - m_V \right)_{\odot} = 0.65 \pm 0.01$ and $\left(m_V - m_R \right)_{\odot} = 0.35 \pm 0.01$ (Ram\'{i}rez et al. 2012). Given the uncertainties, the color of 2016 J1 is indistinguishable from colors of the C- and G-type asteroids (e.g., Dandy et al. 2002).  The latter are believed to have undergone minimal thermal processing.  

We then investigate the normalized reflectivity gradients $S'_{j}\left( \lambda_1, \lambda_2 \right)$ of the two components ($j=\mathrm{a,b}$), defined by Jewitt \& Meech (1986) as

\begin{equation}
S'_{j}\left( \lambda_1, \lambda_2 \right) = \left( \frac{2}{\left| \Delta \lambda_{1,2} \right|} \right) \frac{10^{0.4 \left[ \Delta m_{1,2}^{(j)} - \Delta m_{1,2}^{(\odot)}\right] } - 1}{10^{0.4 \left[ \Delta m_{1,2}^{(j)} - \Delta m_{1,2}^{(\odot)}\right] } + 1}
\label{eq_sp},
\end{equation}

\noindent where $\Delta \lambda_{1,2}$ is the difference in the effective wavelengths of the $BVR$ filter pairs, $\Delta m_{1,2}$ is the color index in the pair, and the superscript in parentheses indicates the corresponding object. For component  J1-A, we obtain $S'_{\mathrm{a}} (V,B) = \left(7.3 \pm 3.6 \right)$\% per 10$^3$ \AA, and $S'_{\mathrm{a}} (V,R) = \left(0.7 \pm 3.9 \right)$\% per 10$^3$ \AA. For component J1-B, we have $S'_{\mathrm{b}} (V,B) = \left(7.5 \pm 10.0 \right)$\% per 10$^3$ \AA, and $S'_{\mathrm{b}} (V,R) = \left(3.6 \pm 11.8 \right)$\% per 10$^3$ \AA. By comparison, the optical continuum reflectivity gradients of the active comets analyzed by Jewitt \& Meech (1986) vary from $\left(5 \pm 2 \right)$\% to $\left(18 \pm 2 \right)$\% per 10$^3$ \AA. Therefore, it is likely that the color of 2016 J1 is amongst the bluest of the measured coma colors, but is similar to those of the active asteroids, e.g., 133P/Elst-Pizarro (Hsieh et al. 2004), and 259P/Garradd (Jewitt et al. 2009). We expect that the color is dominated by scattering from dust particles in the coma, rather than from the nucleus surface.\footnote{The expectation is based on comparison of the FWHM of stars (see Section \ref{sect_obs}) and that of the two components (FWHM $\sim 1\arcsec.3$ for J1-A, and $\sim$1\arcsec.9 for J1-B, respectively). } A possibility that the scattering properties of the nucleus surface significantly differ from those of the dust grains in the coma cannot be ruled out. However, the Keck data do not reveal any statistically significant spatial variations in color.

We  calculate the effective cross sections from the photometry using

\begin{equation}
C = \frac{\pi r_{\mathrm{H}}^2 \mathit{\Delta}^2}{p_\lambda r_\oplus^2 \phi (\alpha)} 10^{-0.4 \left(m_\lambda - m_{\odot,\lambda} \right)}.
\label{eq_xs}
\end{equation}

\noindent Here, $r_{\mathrm{H}}$ and $\mathit{\Delta}$ are  the heliocentric and geocentric distances, both in meters, $r_{\oplus} \approx 1.5\times 10^{11}$ m is the mean Sun-Earth distance, $p_{\lambda}$ is the geometric albedo, and $\phi (\alpha)$ is the phase function of the coma, which we approximate as the empirical phase function of dust (Marcus 2007; Schleicher \& Bair 2011; \url{http://asteroid.lowell.edu/comet/dustphase.html}), and normalize at $\alpha=0$\degr. To avoid potential gas contamination in the comae of components J1-A and J1-B, we focus on $R$-band only. With a nominal, comet-like albedo $p_{R} = 0.04$ (Lamy et al. 2004) for both components of 2016 J1, Equation (\ref{eq_xs}) yields $C = 2.4$ km$^2$ for J1-A, and $0.5$ km$^2$ for J1-B.  Uncertainties in $C$ are dominated by the unmeasured geometric albedo, which could easily be 50\% smaller or larger (c.f.~Lamy et al. 2004).

If  we further assume that the total cross-section within our photometric aperture (radius $\vartheta$ = 9.7$\times 10^{-6}$ radians) is produced in a steady state with some effective particle speed ($v \sim 1$ m s$^{-1}$, see Section \ref{sect_morph}), then the residence time in the aperture is $\vartheta \mathit{\Delta} / v$ and the required  mass-loss rate is 

\begin{equation}
\dot{M} \sim \frac{\rho_{\mathrm{d}} C \bar{\mathfrak{a}} v}{\vartheta \mathit{\Delta}}
\label{eq_dotM},
\end{equation}

\noindent where $\rho_{\mathrm{d}} \sim$ 10$^3$ kg m$^{-3}$ is the bulk density of the dust grains, $v$ is their speed and $\bar{\mathfrak{a}}$ is their average radius. Substituting, $v$ =  1 m s$^{-1}$, $\bar{\mathfrak{a}}$ = 10$^{-3}$ m, we obtain $\dot{M} \sim 1$ kg s$^{-1}$ for J1-A, and $\sim$0.1 kg s$^{-1}$ for J1-B.  For comparison, the mass-loss rates of J1-A and J1-B inferred by extrapolation of an empirical relationship between absolute magnitude and gas production rate (Jorda et al.~2008) are $\dot{M} \approx 1.1$ kg s$^{-1}$ and 0.4 kg s$^{-1}$, respectively, in better agreement with the estimates from the photometry than we could reasonably expect.

The photometry can also be used to provide upper limits to the sizes of  J1-A and J1-B, by assuming that the inferred cross-sections, $C = 2.4$ km$^2$ for J1-A, and $0.5$ km$^2$ for J1-B, are equal to those of the underlying nuclei.  The effective radii of  equal-area circles are calculated from $R_\mathrm{N} \le \sqrt{C / \pi}$, giving $R_\mathrm{N} \approx 0.9$ km and $0.4$ km for J1-A and J1-B, respectively. These are strong upper limits to the true nucleus radii because of the contaminating effects of near-nucleus dust.

Crude lower limits to the nucleus radii of J1-A and J1-B can be placed by assuming that the activity is driven by the equilibrium sublimation of exposed ice. We solved the energy balance equation to calculate the specific sublimation rate of $f_{\mathrm{s}} = 1.6 \times 10^{-5}$ kg m$^{-2}$ s$^{-1}$ at $r_{\mathrm{H}}$ = 2.46 AU. To supply mass loss at the rate of $\sim$1 kg s$^{-1}$ would then require an exposed area $\sim$$6.3 \times 10^{4}$ m$^2$, corresponding to a circular surface patch roughly 140 m in radius, which we take as a lower limit to the nucleus radius of J1-A.  The equivalent lower-limit radius of J1-B is $\sim$40 m. These estimates are clearly very uncertain, because we do not know that the activity is driven by sublimation and, even if it is, we do not know that the sublimation occurs in equilibrium.

In summary, the nucleus radii are weakly constrained by the observations to lie in the range $140 \lesssim R_{\mathrm{N}} \lesssim 900$ m for J1-A, and $40 \lesssim R_{\mathrm{N}} \lesssim 400$ m for J1-B. The photometry suggests that J1-A is larger than J1-B, because J1-A is brighter, but evidence from split comets indeed shows that brightness and nucleus size are often not well-correlated, with small fragments being more active, per unit area, than their larger, parent bodies (Boehnhardt 2004).  As a result, we cannot be sure of the relative sizes of J1-A and J1-B.

\section{\uppercase{Morphology}}
\label{sect_morph}

\begin{figure}
\epsscale{1.0}
\begin{center}
\plotone{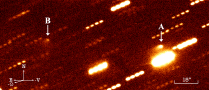}
\caption{
Keck I average image of components J1-A and J1-B (both arrowed and labeled) of active asteroid P/2016 J1 (PANSTARRS) on UT 2016 August 4 coadded from the individual $B$-, $V$- and $R$-band images. As indicated by a compass in the lower left corner, equatorial north is up and east is left. The antisolar direction ($\theta_\odot$) and the negative heliocentric velocity vector projected on the sky plane ($\theta_{\mathbf{V}}$), respectively, are both shown. Also shown is a scale bar. The region shown is $\sim$$2\arcmin.2 \times 1\arcmin.0$. The dotted streaks are trails of background stars and galaxies.
\label{fig_16J1_160804}
} 
\end{center} 
\end{figure}

Components J1-A and J1-B present similar morphologies in our Keck images, although the latter appears much fainter (Figure \ref{fig_16J1_160804}). There is no evidence for a dust trail along the line of the projected orbit, as would be expected of very old, slow, presumably large particles ejected from either nucleus.  To set a zeroth-order constraint on the dust properties implied by the morphology, we first adopt a method based on Finson \& Probstein (1968) to calculate a syndyne-synchrone grid for 2016 J1. In this model, the dust grains are assumed to be released from the surface with zero initial velocity and are then progressively accelerated by solar radiation pressure. The particle trajectory is determined by the release time ($\tau$) and the ratio between the solar radiation pressure force and the local gravitational force due to the Sun ($\beta$). A synchrone  is the locus of positions of grains having varying $\beta$ but with a common release time, and a syndyne is the line marking grains  released at different times but subject to a common $\beta$. 

\begin{figure}
\epsscale{0.8}
\begin{center}
\plotone{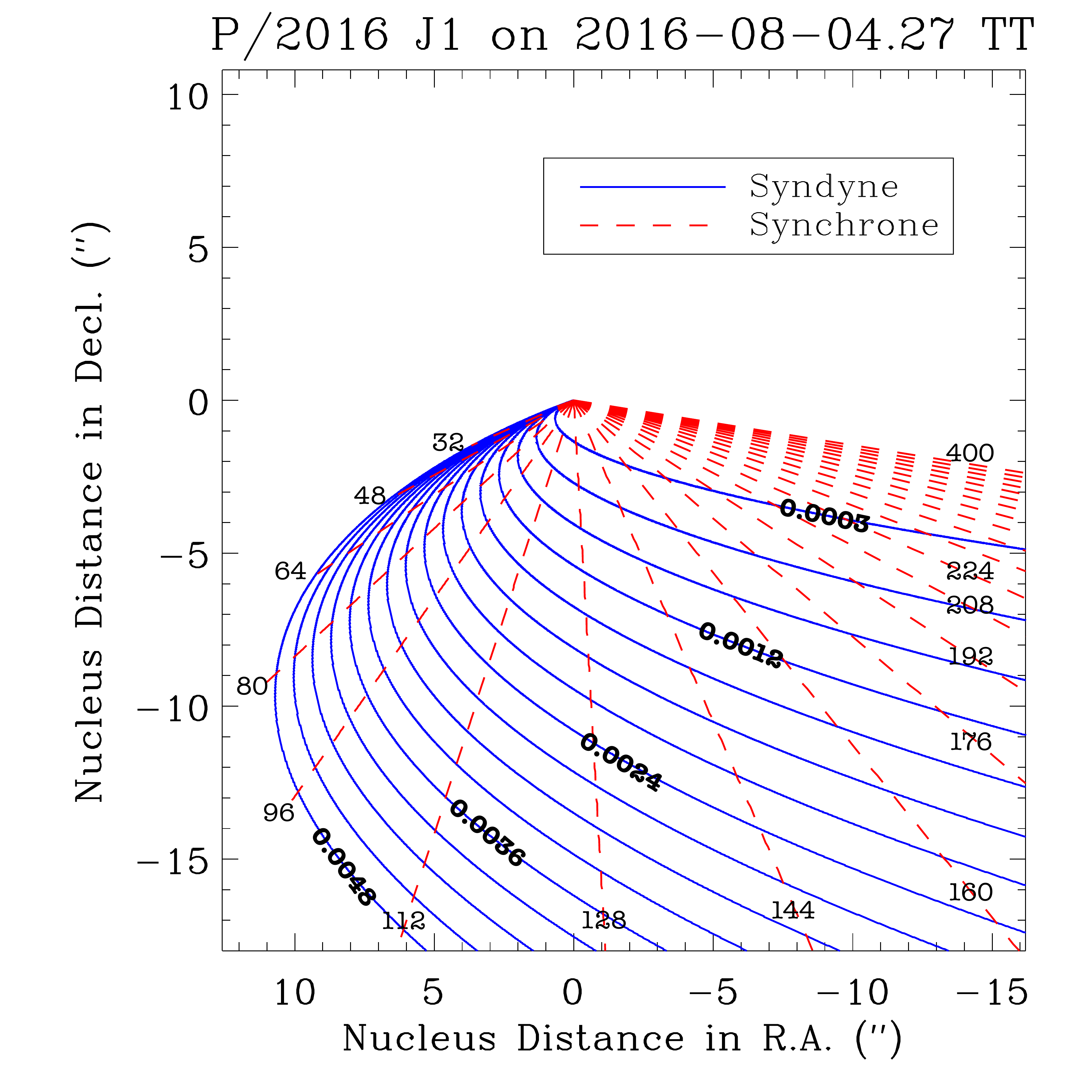}
\caption{
Syndyne-synchrone diagram for P/2016 J1 in comparison to Figure \ref{fig_16J1_160804}. Because of similarities of the orbits of two components, syndyne-synchrone differences are unnoticeable. As indicated, the blue lines are the syndynes, and the synchrones are in dashed red. For clarity some of the syndynes and synchrones are labelled. The synchrone lines correspond to ejections from older days prior to the epoch from left to right, with the leftmost 16 days, rightmost 400 days, and an interval of 16 days. The syndyne lines have $\beta$ decreasing anticlockwise, with the left-most $\beta = 4.8 \times 10^{-3}$, right-most $3 \times 10^{-4}$, and an interval of $3 \times 10^{-4}$.
\label{fig_16J1_fp}
} 
\end{center} 
\end{figure}

The result is plotted in Figure \ref{fig_16J1_fp}. Although J1-B is very faint, we can still trace its fan-shaped tail which is compatible with a series of syndynes having $\beta \lesssim 10^{-3}$ and $\tau \lesssim 200$ days. Interpreting J1-A quantitatively, however, is influenced by a nearby background star. Nevertheless, we are able to identify that the tail of J1-A is produced by $\beta \lesssim 2 \times 10^{-3}$ and $\tau \lesssim 100$ days. The grain size is related to the bulk density, $\rho_{\mathrm{d}}$, and $\beta$ by $\mathfrak{a} \propto \left( \rho_{\mathrm{d}} \beta \right)^{-1}$, from which we can see that 2016 J1 has been releasing dust particles of at least submillimeter-size, given an assumed density $\rho_{\mathrm{d}} \sim$ 10$^3$ kg m$^{-3}$.

The syndyne-synchrone simulation alone does not provide any constraint on the ejection speeds of the dust grains, denoted as $v_{\mathrm{ej}}$. The absence of a strong sunward extent to the coma, however, sets a rough limit to the ejection speed through the following equation

\begin{equation}
v_{\mathrm{ej}} \le \frac{\sqrt{2 \beta G M_{\odot} \mathit{\Delta} \ell }}{r_{\mathrm{H}}}
\label{eq_turnar},
\end{equation}

\noindent where $G = 6.67 \times 10^{-11}$ N kg$^{2}$ m$^{2}$ is the gravitational constant, $M_{\odot} \approx 2 \times 10^{30}$ kg is the mass of the Sun, and $\ell$ is the apparent sunward turnaround distance expressed in radians. Our Keck images show $\ell \approx 1\arcsec.0$ for both components. By substitution with the corresponding values we obtain $v_{\mathrm{ej}} \approx 2.4$ and 1.7 m s$^{-1}$ from Equation (\ref{eq_turnar}) for submillimeter-sized dust grains of components J1-A and J1-B, respectively. The low ejection speed is similar to speeds measured in some other active asteroids, such as 133P/(7968) Elst-Pizarro (Jewitt et al. 2014b) and 313P/La Sagra (Jewitt et al. 2015b), in which little or no sunward extent of the coma is observed.  The likely cause is weak gas flow from a small source, or perhaps through a porous mantle.

We also considered a more realistic model where dust grains are emitted from the nucleus with non-zero initial velocities in a nucleus-centric frame, symmetric about the heliocentric radial direction facing towards the Sun, and satisfying a power-law size distribution. Similar to Ishiguro (2008), the ejection terminal speed is set to be correlated with the dust size and heliocentric distance, empirically expressed as

\begin{equation}
v_{\mathrm{ej}} = v_{0} \sqrt{ \frac{\mathfrak{a}_{0} r_{\oplus}}{\mathfrak{a} r_{\mathrm{H}}} }
\label{eq_Vej},
\end{equation}

\noindent where $v_{0}$ is the ejection speed of dust particles of $\mathfrak{a}_{0} = 0.5$ cm in radius at heliocentric distance $r_{\mathrm{H}} = 1$ AU. The positions of generated dust grains at the observed time are then obtained by means of orbital integration in our modified version of the \textit{mercury6} package (Chambers 1999), with inclusion of perturbations due to the eight major planets, Pluto, and the most ten massive main-belt asteroids, although these perturbation effects are found to be minimal. Finally, we calculate a model image in which the intensity of each pixel is $\mathcal{I}(x,y)$, 

\begin{equation}
\mathcal{I} \left( x, y \right) \propto \iint \left( \frac{\mathfrak{a}}{r_{\mathrm{H}}} \right)^{2} \mathfrak{D} \left(x', y' \right) \mathrm{d}x' \mathrm{d}y' .
\label{eq_flux}
\end{equation}

\noindent Here $\mathfrak{D}$ is the surface density of the dust particles in the sky plane as a function of pixel coordinates. In theory, a set of best-fit parameters can be solved such that the modelled image matches the observation the most closely. However, limitations of the data prevent us from obtaining meaningful results in the most general case. Instead, we vary only $v_{0}$ in Equation (\ref{eq_Vej}), whilst the differential power-law index $\mathit{\gamma}$ remains fixed, because this is the parameter that will most affect the modelled morphology. 

\begin{figure}
\epsscale{1.0}
\begin{center}
\plotone{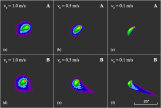}
\caption{
Modelled morphology of components J1-A (the first row, as indicated by the letter at the upper right corner of each panel) and J1-B (the second row) of active asteroid P/2016 J1 as a function of $v_{0}$. The values of $v_{0}$ are labelled explicitly for each panel. Other parameters remain fixed in the simulations for the same objects, i.e., the grain radius $0.3 \le \mathfrak{a} \le 100$ mm and the released time $\tau \le 100$ days for component  J1-A, and $0.6 \le \mathfrak{a} \le 100$ mm and $\tau \le 200$ days for component B. All the simulations have the differential slope index of the dust-size distribution $\mathit{\gamma} = 3.5$. A number of $\sim10^{4}$ particles are generated in each run of the simulation. The modelled images are smoothed to mimic the seeing during the Keck observation. Also given is a scale bar, which is applicable to all the panels. Equatorial north is up, and east is left. By visual comparison, we conclude that the modelled images from the middle column best match the data.
\label{fig_16J1_model}
} 
\end{center} 
\end{figure}

By trial and error, we determine that when $v_{0} \sim 0.5$ m s$^{-1}$, the modelled and the observed morphology match the best, (Figure \ref{fig_16J1_model}), a result which is true for both components. Larger speeds, e.g., $v_{0} = 1$ m s$^{-1}$,  create a tail that is too wide compared to the actual tail, while smaller speeds, e.g., $v_{0} = 0.1$ m s$^{-1}$, result in tails that are too narrow. This conclusion is broadly consistent with our estimate of the ejection speed of submillimeter-sized dust grains as a few meter per second, provided that Equation (\ref{eq_Vej}) is valid. Other parameters do affect the morphology, but a detailed and exhaustive investigation is beyond the scope of this paper.

\section{\uppercase{Split Dynamics}}
\label{sect_sd}

The similarities in the orbits of J1-A and J1-B of 2016 J1 (see Table \ref{tab_orb}) indicate a common origin. To investigate when the two components were produced, a straightforward method is to perform orbital integration backward for both components. However, we found that uncertainties in astrometric measurements have prevented us from obtaining a good constraint on the split epoch, because this method at best yields a nonzero close encounter distance $\sim$10$^{-4}$ AU between the two components, which is unrealistic. We therefore created clones for each of the components based on the covariance matrices of the orbital elements and performed orbital integration for every clone. We then monitored their evolution in three-dimensional space visually in terms of orthogonal equatorial coordinates J2000, and observed that the two ellipsoids of clone clouds overlap with each other twice. The most recent period of overlap occurred $\sim$1500--1800 days prior to the epoch of observation (roughly from 2011 November to 2012 April), and the earlier one occurred $\sim$2500 days ago (2010 March to May).\footnote{This part of work was done using the \textit{SOLEX} package by A. Vitagliano.}

A potential nongravitational force will delay the best overlap epoch. Since the observing arc of 2016 J1 is too short, solving nongravitational parameters based on the astrometry only yields an unreliable result. Indeed, we  have no detection. Besides, this method cannot provide good constraints on other parameters, such as the relative separation velocity of the secondary object. Dissatisfied, we exploited a different approach where we search for the best-fit fragmentation parameters, including the split epoch $t_\mathrm{frg}$, the dimensionless radial nongravitational parameter of the secondary object $\tilde{\beta}$, which is mathematically the same as $\beta$ but which arises from anisotropic sublimation of volatiles, and the three Cartesian components of the separation velocity $\mathbf{V}_\mathrm{r}$, $\mathbf{V}_\mathrm{t}$, and $\mathbf{V}_\mathrm{n}$, in the radial, transverse, and normal directions instantly at $t_\mathrm{frg}$ centered at the primary object. A similar technique was developed by Sekanina (1978, 1982),  and has been applied repeatedly (e.g., Meech et al.~1995; Sekanina \& Chodas 2002). A major difference exists in terms of dealing with the nongravitational acceleration between the two models, in that the model by Sekanina (1978, 1982) solves for the differential deceleration of the secondary component with respect to the primary. In cases where the primary nucleus has no detectable nongravitational effects, the difference disappears.

At the splitting epoch $t_\mathrm{frg}$, position coordinates of the primary nucleus and the fragment are  the same, only with a difference in their velocities. Gravitational interactions between the components are ignored. We then integrate the state vector of the secondary by the \textit{mercury6} package in combination with planetary ephemeris DE431 and perturbations from the eight major planets, Pluto, and the most ten massive main-belt asteroids all included. Topocentric positions of the secondary at each observed epoch are then compared to the observed position. We iterate the same procedure until the split parameters minimize the following quantity

\begin{equation}
\chi^2 \left(t_\mathrm{frg}, \mathbf{V}_\mathrm{r}, \mathbf{V}_\mathrm{t}, \mathbf{V}_\mathrm{n}, \tilde{\beta} \right) = \sum_{i=1}^{N} \mathcal{R}_i^2 w_i^2,
\label{eq_chi2}
\end{equation} 

\noindent where $N$ is the total number of scalar astrometric observations, $\mathcal{R}_i^2$ is the squared astrometric residual in Right Ascension (R.A.) and declination (decl.), and $w_i$ is the weight of the $i$-th observation assigned based on the observation quality of components J1-A and J1-B.\footnote{The task was accomplished by exploiting \textit{MPFIT} (Markwardt 2009).} Here, different astrometric observations are assumed to be uncorrelated with each other, which is usually a good approximation. 

We first regarded J1-B as the principal nucleus of the active asteroid, because it is the leading component, and was apparently located close to the negative heliocentric velocity vector projected onto the sky plane. In the process of finding the best-fit solution, we attempted various approaches. First, we searched for a solution mathematically equivalent to a syndyne-synchrone computation, where the total initial separation velocity $\mathbf{V}_{\mathrm{sep}}$ was forced to be zero, whilst $t_{\mathrm{frg}}$ and $\tilde{\beta}$ were the two free parameters to be optimised, with an initial guess of $t_{\mathrm{frg}}$ in 2012. The solution did converge, but the result cannot be accepted, for it produced a strong systematic bias in the  astrometric residuals considerably greater than the assigned uncertainties, along with a huge $\chi^2 > 2.1 \times 10^{4}$. Second, we included the three components of $\mathbf{V}_{\mathrm{sep}}$ as free parameters to be solved, resulting in $\chi^2$ declining appreciably and finally converging to $\chi^2 = 17.7$. However, this solution produced $\tilde{\beta} < 0$ with a signal-to-noise ratio (SNR) $< 5$. We regard this as likely bogus because a single run heavily relies on the orbit of J1-B, the quality of which is not optimum. To address this, we created 100 Monte Carlo (MC) clones of J1-B in accord with the covariance matrix of its orbital elements. For each of the clones, a set of best-fit split parameters were optimised. In this way we found a less significant $\tilde{\beta}$ (SNR = 3.3). 

We then forced $\tilde{\beta} \equiv 0$ exactly, and exploited the routine to search for four other free parameters ($t_{\mathrm{frg}}$ and the three Cartesian components of $\mathbf{V}_{\mathrm{sep}}$). Although the converged solution yielded a larger $\chi^{2} = 35.0$, we did not notice any systematic bias beyond or comparable to the uncertainty levels. The same procedures were attempted for finding a solution with an initial guess of $t_{\mathrm{frg}}$ in 2010. Our routine yielded an equally good solution (e.g., $\chi^{2} = 17.7$ for all the five splitting parameters regarded as free parameters producing $\tilde{\beta} < 0$, and $\chi^{2} = 22.2$ for the scenario of $\tilde{\beta} \equiv 0$). We therefore conclude that there is no compelling evidence for the existence of nongravitational effects on component J1-A, consistent with the aforementioned fact that the ordinary orbit determination revealed no statistically significant nongravitational parameters. Instead, the relative positions of J1-A and J1-B are overwhelmingly determined by the separation velocity of the two fragments, rather than by any differential non-gravitational force. For this reason, we cannot use dynamics to ambiguously determine which component is the principal nucleus of 2016 J1.

We then took advantage of the fact that the orbital solution of J1-A has a better quality and solved the splitting parameters by regarding J1-A as the primary. Again there is no reliable detection of $\tilde{\beta}$ for component J1-B, and so we set $\tilde{\beta} \equiv 0$. Table \ref{tab_split} summarizes the two equally good best-fit solutions for $t_{\mathrm{frg}}$ in 2012 and 2010, and we present the corresponding astrometric residuals in Table \ref{tab_res}. We also created 100 MC clones of J1-A from its covariance matrix of the orbital elements, and performed optimization repeatedly. The statistics are listed in Table \ref{tab_split}, consistent with the results from the single runs. The split event either occurred at some point around 2012 May to June, when 2016 J1 was $r_{\mathrm{H}} \approx 3.4$--3.5 AU from the Sun, with a total separation speed $V_\mathrm{sep} = 0.70 \pm 0.02$ m s$^{-1}$, or in 2010 April, when it was $r_{\mathrm{H}} \approx 2.6$ AU, with a total separation speed $V_\mathrm{sep} = 0.83 \pm 0.06$ m s$^{-1}$, mainly in the radial direction.  We have no strong basis for preferring one solution over the other.  The separation speed value falls within the range of values shown by split Kuiper belt and Oort cloud comets (c.f.~Sekanina 1982, Boehnhardt 2004). Unfortunately, the physical mechanism leading to fragmentation of 2016 J1 cannot be constrained by dynamics, because different splitting mechanisms can lead to  similar separation speeds.

\subsection{\uppercase{Lifetime}}
For illustrative purposes in this section, we adopt a radius $R_{\mathrm{N}} \sim$ 300 m for both components of 2016 J1; this value is consistent with the observational constraints described above.  We evaluated the de-volatilization timescale $\tau_{\mathrm{dv}}$ by

\begin{equation}
\tau_{\mathrm{dv}} \sim \frac{\rho_{\mathrm{d}} R_{\mathrm{N}}}{ f_{\mathrm{s}}}
\label{eq_tdv},
\end{equation}

\noindent where $f_{\mathrm{s}}$ is the time-averaged equilibrium specific sublimation  rate around the orbit. In its present orbit, 2016 J1 travels from $r_{\mathrm{H}}$ = 2.45 AU at perihelion to 3.89 AU at aphelion. Over this range, we calculate $f_{\mathrm{s}} = 4.4\times 10^{-6}$ kg m$^{-2}$ s$^{-1}$.  Substituting  the values from Section \ref{sect_phot}, Equation (\ref{eq_tdv}) yields $\tau_{\mathrm{dv}} \sim 7 \times 10^{10}$ s, or $\sim$2 kyr as the approximate de-volatilization timescale for 2016 J1. It is worth pointing out that this value is only a lower limit, because the growth of a refractory mantle is likely to impede outgassing as a result of an aging nucleus (Rickman et al. 1990). 

A short lifetime is also indicated by simple models of rotational instability, driven by outgassing torques (Jewitt 1997, Jewitt et al.~2016). For example, Equation (3) of the latter paper gives an $e$-folding spin-up time $\tau_{\mathrm{s}} \sim 0.2$ kyr, for an assumed period of 5 hr, radius 0.3 km, outgassing speed $\sim$0.5 km s$^{-1}$, dimensionless moment arm of the torque $k_{\mathrm{T}} = 10^{-3}$, and mass loss rate $\dot{M} \sim 1$ kg s$^{-1}$. The moment arm is unknown to within at least an order of magnitude ($0.0004 \le k_{\mathrm{T}} \le 0.04$; Guti\'{e}rrez et al. 2003; Belton et al. 2011; Drahus et al. 2011) so that this timescale, like $\tau_{\mathrm{dv}}$, is approximate. Nevertheless, it is evident that the components of 2016 J1 are likely to be short-lived.

Given that the sublimation and rotational spin-up timescales are small compared to the 4.5 Gyr age of the solar system, the question arises as to why 2016 J1 survives at all. The answer, as is the case for the main-belt comets generally, presumably lies with the long-term history and stability of the near-surface ice (e.g., Hsieh \& Jewitt 2006). Conduction models show that near-surface ice can be stabilized against sublimation by a modest (meter-sized thick) refractory, particulate layer (Schorghofer 2008). Buried ice can persist in the asteroid belt with negligible sublimation losses for times comparable to the age of the solar system.  Only when exposed at the surface, by a small impact or other disturbance, can the ice sublimate.  We thus imagine that the long-term survival of  asteroid 2016 J1 reflects the past burial and recent excavation of its near-surface ice. In this scenario, a small, surface-disturbing impact can expose ice which produces a sublimation torque leading to break-up of the nucleus.  Further splitting events are possible in the near future. We hence encourage followup observations of 2016 J1.

\clearpage

\section{\uppercase{Summary}}
Key conclusions of our analysis of active asteroid P/2016 J1 (PANSTARRS) are summarized as follows.

\begin{enumerate}

\item The radii of the components are constrained to lie in the range $140 \lesssim R_{\mathrm{N}} \lesssim 900$ m, for J1-A and $40 \lesssim R_{\mathrm{N}} \lesssim 400$ m, for J1-B, assuming a comet-like geometric albedo $p_{R} = 0.04$.

\item The separation between J1-A and J1-B occurred either $\sim$1500 days (around 2012 May to June) or 2300 days (in 2010 April) before the present observations. The separation speed is $V_{\mathrm{sep}} = 0.70 \pm 0.02$ m s$^{-1}$ for the former case, and $0.83 \pm 0.06$ m s$^{-1}$ for the latter, mainly in the radial direction. A dust dynamics simulation shows that both components have been active (at rates up to $\sim$1 kg s$^{-1}$) for several months, but not for the full $\sim$1500 or 2300 days since break-up.

\item The component colors ($m_{B} - m_{V} = 0.74 \pm 0.04$ and $m_{V} - m_{R} = 0.36 \pm 0.04$ for J1-A, and $m_{B} - m_{V} = 0.74 \pm 0.12$ and $m_{V} - m_{R} = 0.39 \pm 0.12$ for J1-B), are the same within the uncertainties. These nearly neutral colors are consistent with the spectra of primitive C- and G-type asteroids.

\item The sublimation and rotational spin-up lifetimes of J1-A and J1-B are much shorter than the 4.5 Gyr age of the solar system, implying that the observed activity cannot be sustained. 

\item The breakup of 2016 J1 may itself be due to rotational instability induced by sublimation torques in an asteroid having recently-exposed  surface ice. We suggest that further disintegration events are possible in the near future, owing to continued rotational instability.

\end{enumerate}

\acknowledgements
We thank Marco Micheli for his astrometric measurements in our Keck-I images and Davide Farnocchia for updating the orbits of the two components in JPL HORIZONS. We also thank observers who submitted astrometry of 2016 J1 to the MPC, from which our study has benefited. We further appreciate comments from an anonymous referee on this manuscript. MTH appreciates discussions with Quan-Zhi Ye. The data presented herein were obtained at the W. M. Keck Observatory, which is operated as a scientific partnership among the California Institute of Technology, the University of California, and the National Aeronautics and Space Administration. The Observatory was made possible by the generous financial support of the W. M. Keck Foundation. This work is funded by a grant from NASA's Solar System Observations program to DJ.

\begin{deluxetable}{l|cc|cc}
\rotate
\tablecaption{
Orbital Solutions of Pair P/2016 J1-A and -B
\label{tab_orb}
}
\tablewidth{0pt}
\tablehead{ Orbital Element & 
\multicolumn{2}{c}{Component J1-A}  & 
\multicolumn{2}{c}{Component J1-B}  \\
 & 
\colhead{Value} & \colhead{1$\sigma$ Uncertainty} & 
\colhead{Value} & \colhead{1$\sigma$ Uncertainty} 
}
\startdata
Semimajor axis $a$ (AU) 
       & 3.172092 & 1.83$\times$10$^{-5}$ %
       & 3.172014 & 2.18$\times$10$^{-5}$ \\ %
Perihelion distance $q$ (AU) 
       & 2.448015 & 1.25$\times$10$^{-5}$ %
       & 2.448063 & 1.57$\times$10$^{-5}$ \\ %
Eccentricity $e$ 
       & 0.228265 & 3.45$\times$10$^{-6}$ %
       & 0.228231 & 5.42$\times$10$^{-6}$ \\ %
Inclination $i$ (\degr) 
       & 14.330194 & 9.59$\times$10$^{-5}$ %
       & 14.331156 & 1.35$\times$10$^{-4}$ \\ %
Longitude of ascending node $\Omega$ (\degr)
                 & 199.856231 & 2.15$\times$10$^{-4}$ %
                 & 199.855506 & 3.28$\times$10$^{-4}$ \\ %
Argument of perihelion $\omega$ (\degr)
                 & 46.585537 & 4.50$\times$10$^{-3}$ %
                 & 46.579240 & 6.68$\times$10$^{-3}$ \\ %
Time of perihelion $t_\mathrm{P}$ (TT)
                 & 2016 Jun 24.2138 & 1.49$\times$10$^{-2}$ %
                 & 2016 Jun 24.1244 & 2.21$\times$10$^{-2}$ \\ %

\enddata
\tablecomments{
The orbital solutions were taken from the JPL Small-Body Database Browser. Osculating epochs for components J1-A and J1-B are TT 2016 Jun 20.0 and May 31.0, respectively. Normalized rms, which is defined as $\pm \sqrt{\sum_{i=1}^{N} \mathcal{R}_i^2 w_i^2 / N}$, for component J1-A is $\pm$0.392 from 66 observations spanning UT 2016 March 04 -- August 04, and for J1-B is $\pm$0.455 from 51 observations spanning UT 2016 March 17 -- August 04.
}
\end{deluxetable}

\begin{deluxetable}{l|cc|cc}
\tabletypesize{\footnotesize}
\rotate
\tablecaption{
Fragmentation Solutions for Pair P/2016 J1-A and -B
\label{tab_split}
}
\tablewidth{0pt}
\tablehead{
Quantity & 
\multicolumn{2}{c}{Solution I}  & \multicolumn{2}{c}{Solution II}  \\
& \colhead{Single Run} & \colhead{MC Iteration\tablenotemark{\dagger}} 
& \colhead{Single Run} & \colhead{MC Iteration}
}
\startdata
Split epoch $t_\mathrm{frg}$ (TT) & 2012 May $18.6 \pm 14.8$ & 2012 May $19.5 \pm 8.1$ & 
                                                          2010 Apr $11.8 \pm 8.3$ & 2010 Apr $08.7 \pm 3.5$ \\ %
Separation velocity (m s$^{-1}$): & & & & \\
~~~~Radial component $\mathbf{V}_\mathrm{R}$ 
& $-0.6450 \pm 0.0169$ & $-0.6458 \pm 0.0079$ 
& $+0.7486 \pm 0.0597$ & $+0.7556 \pm 0.0258$ \\ %
~~~~Transverse component $\mathbf{V}_\mathrm{T} $ 
& $ +0.0295 \pm 0.0063$ & $ +0.0296 \pm 0.0029$
& $ -0.0327 \pm 0.0089$ & $ -0.0309 \pm 0.0035$ \\
~~~~Normal component $\mathbf{V}_\mathrm{N}$ 
& $-0.2798 \pm 0.0089$ & $-0.2803 \pm 0.0040$ 
& $+0.3624 \pm 0.0180$ & $+0.3553 \pm 0.0064$ \\
Weighted rms\tablenotemark{\ast} (\arcsec) 
& $\pm$0.123 & $\in \pm \left[ 0.105, 0.154 \right] $ 
& $\pm$0.101 & $\in \pm \left[ 0.099, 0.126 \right] $ \\
Normalised rms\tablenotemark 
& $\pm$0.569 & $\in \pm \left[0.483, 0.713 \right] $
& $\pm$0.465 & $\in \pm \left[0.459, 0.580 \right] $ \\

\enddata
\tablenotetext{\dagger}{~100 clones of synthesised J1-A are created based on the covariance matrix of its orbital elements. The uncertainties are standard deviations on the mean values.}
\tablenotetext{\ast}{~The weighted rms is calculated from $\pm \sqrt{\sum_{i=1}^{N} \mathcal{R}_i^2 w_i^2 / \sum_{i=1}^{N}w_i^2}$. For the MC iterations, the minimum and the maximum of the unsigned weighted rms are given.}
\tablecomments{
Both methods completely selected the same astrometric data and adopted the same weighting scheme as those used to solve an orbit for component J1-B in Table \ref{tab_orb}. The dimensionless nongravitational parameter $\tilde{\beta}$ is forced to be zero in all of the runs.
}
\end{deluxetable}

\begin{deluxetable}{lcccccc}
\tablecaption{
Astrometric Residuals in Best-Fit Solutions for P/2016 J1-A and -B
\label{tab_res}
}
\tablewidth{0pt}
\tablehead{
\colhead{Observation Time} &  
\multicolumn{2}{c}{Residuals I\tablenotemark{\dagger} (\arcsec)}  &
\multicolumn{2}{c}{Residuals II\tablenotemark{\ddagger} (\arcsec)}  
& \colhead{Uncertainty}  & \colhead{Observatory} \\
\colhead{(UT)} & \colhead{R.A.} & \colhead{Decl.} & \colhead{R.A.} & \colhead{Decl.} 
& \colhead{(\arcsec)} & \colhead{Code\tablenotemark{\ast}}
}
\startdata

2016 Mar 17.622098 & $+0.268$ & $-0.269$ & $+0.097$ & $+0.052$ & 0.30 & 568 \\
2016 Apr 14.53944 & $+0.256$ & $-0.582$ & $+0.047$ & $-0.392$ & 0.70 & F51 \\
2016 Apr 14.55134 & $-0.488$ & $+0.275$ & $-0.697$ & $+0.465$ & 0.70 & F51 \\
2016 Apr 14.56324 & $-0.032$ & $-0.773$ & $-0.241$ & $-0.584$ & 0.70 & F51 \\
2016 Apr 14.57510 & $+0.262$ & $-0.205$ & $+0.053$ & $-0.015$ & 0.70 & F51 \\
2016 May 06.425891 & $+0.137$ & $+0.029$ & $+0.043$ & $+0.085$ & 0.20 & 568 \\
2016 May 06.427167 & $+0.116$ & $+0.037$ & $+0.023$ & $+0.093$ & 0.20 & 568 \\
2016 May 07.445034 & $+0.096$ & $+0.075$ & $+0.010$ & $+0.125$ & 0.20 & 568 \\
2016 May 07.447008 & $+0.160$ & $-0.040$ & $+0.074$ & $+0.010$ & 0.20 & 568 \\
2016 May 07.448979 & $+0.162$ & $+0.017$ & $+0.076$ & $+0.067$ & 0.20 & 568 \\
2016 May 08.349285 & $-0.025$ & $-0.003$ & $-0.104$ & $+0.042$ & 0.50 & H01 \\
2016 May 08.360538 & $+0.057$ & $-0.073$ & $-0.022$ & $-0.028$ & 0.50 & H01 \\
2016 May 08.363089 & $+0.139$ & $+0.083$ & $+0.060$ & $+0.128$ & 0.50 & H01 \\
2016 May 08.368020 & $+0.197$ & $-0.309$ & $+0.118$ & $-0.264$ & 0.50 & H01 \\
2016 May 08.371138 & $+0.180$ & $-0.375$ & $+0.101$ & $-0.330$ & 0.50 & H01 \\
2016 May 11.465427 & $+0.087$ & $+0.052$ & $+0.030$ & $+0.080$ & 0.20 & 568 \\
2016 May 11.466126 & $+0.088$ & $-0.240$ & $+0.031$ & $-0.212$ & 0.30 & 568 \\
2016 May 11.469146 & $+0.161$ & $-0.050$ & $+0.105$ & $-0.022$ & 0.15 & 568 \\
2016 May 11.471359 & $+0.044$ & $-0.060$ & $-0.013$ & $-0.032$ & 0.15 & 568 \\
2016 May 15.092581 & $+0.061$ & $-0.079$ & $+0.030$ & $-0.069$ & 0.20 & Z18 \\
2016 May 15.094942 & $+0.045$ & $-0.033$ & $+0.014$ & $-0.023$ & 0.20 & Z18 \\
2016 May 15.097292 & $+0.157$ & $-0.052$ & $+0.126$ & $-0.042$ & 0.20 & Z18 \\
2016 May 15.099653 & $+0.201$ & $-0.196$ & $+0.169$ & $-0.186$ & 0.20 & Z18 \\
2016 May 15.102002 & $+0.116$ & $-0.195$ & $+0.085$ & $-0.185$ & 0.20 & Z18 \\
2016 May 29.044549 & $-0.070$ & $+0.062$ & $-0.015$ & $+0.023$ & 0.10 & Z18 \\
2016 May 29.046898 & $-0.092$ & $+0.044$ & $-0.037$ & $+0.006$ & 0.10 & Z18 \\
2016 May 29.049248 & $-0.084$ & $+0.057$ & $-0.029$ & $+0.019$ & 0.10 & Z18 \\
2016 May 29.051609 & $-0.114$ & $+0.076$ & $-0.060$ & $+0.037$ & 0.10 & Z18 \\
2016 May 29.053958 & $-0.123$ & $+0.089$ & $-0.068$ & $+0.050$ & 0.10 & Z18 \\
2016 Jun 06.339455 & $-0.026$ & $-0.072$ & $+0.063$ & $-0.122$ & 0.50 & 695 \\
2016 Jun 07.314874 & $+0.100$ & $-0.403$ & $+0.193$ & $-0.454$ & 0.50 & 695 \\
2016 Jun 09.279018 & $-0.405$ & $-0.011$ & $-0.307$ & $-0.064$ & 0.50 & 695 \\
2016 Jun 11.281689 & $-0.242$ & $-0.128$ & $-0.139$ & $-0.181$ & 0.50 & 695 \\
2016 Jun 12.185745 & $+0.170$ & $-0.349$ & $+0.275$ & $-0.402$ & 0.50 & 695 \\
2016 Jul 31.944329 & $+0.351$ & $-0.243$ & $+0.372$ & $-0.208$ & 0.30 & Z18 \\
2016 Jul 31.946678 & $+0.224$ & $-0.241$ & $+0.245$ & $-0.206$ & 0.30 & Z18 \\
2016 Jul 31.949039 & $+0.149$ & $+0.065$ & $+0.169$ & $+0.100$ & 0.30 & Z18 \\
2016 Jul 31.953738 & $+0.146$ & $-0.231$ & $+0.166$ & $-0.196$ & 0.30 & Z18 \\
2016 Aug 04.269988 & $-0.024$ & $-0.012$ & $-0.015$ & $+0.029$ & 0.30 & 568 \\
2016 Aug 04.270127 & $+0.041$ & $+0.032$ & $+0.050$ & $+0.074$ & 0.30 & 568 \\
2016 Aug 04.274039 & $-0.211$ & $+0.098$ & $-0.202$ & $+0.140$ & 0.30 & 568 \\
2016 Aug 04.274178 & $-0.236$ & $-0.007$ & $-0.227$ & $+0.035$ & 0.30 & 568 \\
2016 Aug 04.278345 & $-0.178$ & $+0.011$ & $-0.169$ & $+0.052$ & 0.30 & 568 \\
2016 Aug 04.278472 & $-0.120$ & $+0.012$ & $-0.111$ & $+0.053$ & 0.30 & 568 \\
2016 Aug 04.282234 & $-0.071$ & $-0.001$ & $-0.062$ & $+0.041$ & 0.30 & 568 \\
2016 Aug 04.282373 & $-0.156$ & $+0.004$ & $-0.148$ & $+0.046$ & 0.30 & 568 \\
2016 Aug 04.287257 & $+0.072$ & $+0.052$ & $+0.080$ & $+0.093$ & 0.50 & 568 \\
2016 Aug 04.287396 & $+0.391$ & $-0.074$ & $+0.400$ & $-0.032$ & 0.50 & 568 \\
2016 Aug 04.292500 & $+0.323$ & $-0.055$ & $+0.332$ & $-0.014$ & 0.50 & 568 \\
2016 Aug 04.297060 & $-0.404$ & $+0.038$ & $-0.396$ & $+0.079$ & 0.50 & 568 \\
2016 Aug 04.297199 & $+0.155$ & $-0.177$ & $+0.164$ & $-0.136$ & 0.50 & 568 \\

\enddata
\tablenotetext{\dagger}{~O$-$C residuals from the single-run best-fit solution I in Table \ref{tab_split}.}
\tablenotetext{\ddagger}{~O$-$C residuals from the single-run best-fit solution II in Table \ref{tab_split}.}
\tablenotetext{\ast}{~Corresponding names of the observatories and their geocentric coordinates are given by \url{http://www.minorplanetcenter.net/iau/lists/ObsCodes.html}.}
\end{deluxetable}

\end{document}